\documentclass[12pt]{iopart}

\usepackage[dvips]{color}
\usepackage{lineno}
\usepackage{iopams}
\usepackage{graphicx}

\begin{document}

\title[Single-electron charge signals in the XENON100 experiment]{Observation and applications of single-electron charge signals in the XENON100 experiment}


\newcommand{\columbia}{\address{$^{1}$ Physics Department, Columbia University, New York, NY, USA}}
\newcommand{\nikhef}{\address{$^{2}$ Nikhef  and the University of Amsterdam, Science Park, Amsterdam, Netherlands}}
\newcommand{\losangeles}{\address{$^{3}$ Physics \& Astronomy Department, University of California, Los Angeles, CA, USA}}
\newcommand{\assergi}{\address{$^{4}$ INFN, Laboratori Nazionali del Gran Sasso, Assergi (AQ), Italy}}
\newcommand{\coimbra}{\address{$^{5}$ Department of Physics, University of Coimbra, Coimbra, Portugal}}
\newcommand{\zurich}{\address{$^{6}$ Physics Institute, University of Z\"{u}rich, Z\"{u}rich, Switzerland}}
\newcommand{\mainz}{\address{$^{7}$ Institut f\"ur Physik \& Exzellenzcluster PRISMA, Johannes Gutenberg-Universit\"at Mainz, Mainz, Germany}}
\newcommand{\weizmann}{\address{$^{8}$ Department of Particle Physics and Astrophysics, Weizmann Institute of Science, Rehovot, Israel}}
\newcommand{\munster}{\address{$^{9}$ Institut f\"ur Kernphysik, Wilhelms-Universit\"at M\"unster, M\"unster, Germany}}
\newcommand{\purdue}{\address{$^{10}$ Department of Physics, Purdue University, West Lafayette, IN, USA}}
\newcommand{\subatech}{\address{$^{11}$ SUBATECH, Ecole des Mines de Nantes, CNRS/In2p3, Universit\'e de Nantes, Nantes, France}}
\newcommand{\torino}{\address{$^{12}$ INFN-Torino and Osservatorio Astrofisico di Torino, Torino, Italy}}
\newcommand{\shanghai}{\address{$^{13}$ Department of Physics \& Astronomy, Shanghai Jiao Tong University, Shanghai, China}}
\newcommand{\bologna}{\address{$^{14}$ University of Bologna and INFN-Bologna, Bologna, Italy}}
\newcommand{\heidelberg}{\address{$^{15}$ Max-Planck-Institut f\"ur Kernphysik, Heidelberg, Germany}}
\newcommand{\houston}{\address{$^{16}$ Department of Physics and Astronomy, Rice University, Houston, TX, USA}}
\newcommand{\bern}{\address{$^{17}$ Albert Einstein Center for Fundamental Physics, University of Bern, Bern, Switzerland}}

\author{E.~Aprile$^{1}$, M.~Alfonsi$^{2}$, K.~Arisaka$^{3}$, 
F.~Arneodo$^{4}$\footnote{Present address: New York University in Abu Dhabi, UAE},
C.~Balan$^{5}$, L.~Baudis$^{6}$,
B.~Bauermeister$^{7}$, A.~Behrens$^{6}$, P.~Beltrame$^{8,3}$, K.~Bokeloh$^{9}$, A.~Brown$^{10}$, E.~Brown$^{9}$,
S.~Bruenner$^{15}$, G.~Bruno$^{4}$, R.~Budnik$^{1}$, J.~M.~R.~Cardoso$^{5}$, W.-T.~Chen$^{11}$, B.~Choi$^{1}$,
A.~P.~Colijn$^{2}$, H.~Contreras$^{1}$, J.~P.~Cussonneau$^{11}$, M.~P.~Decowski$^{2}$, E.~Duchovni$^{8}$,
S.~Fattori$^{7}$, A.~D.~Ferella$^{4,6}$, W.~Fulgione$^{12}$, F.~Gao$^{13}$, M.~Garbini$^{14}$, C.~Ghag$^{3}$,
K.-L.~Giboni$^{1}$, L.~W.~Goetzke$^{1}$, C.~Grignon$^{7}$, E.~Gross$^{8}$, W.~Hampel$^{15}$, R.~Itay$^{8}$,
F.~Kaether$^{15}$, G.~Kessler$^{6}$, A.~Kish$^{6}$, 
J.~Lamblin$^{11}$\footnote{Present address: LPSC, Universit\'e Joseph Fourier, CNRS/IN2P3, INPG, Grenoble, France}, 
H.~Landsman$^{8}$, R.~F.~Lang$^{10}$,
M.~Le~Calloch$^{11}$, C.~Levy$^{9}$, K.~E.~Lim$^{1}$, Q.~Lin$^{13}$, S.~Lindemann$^{15}$, M.~Lindner$^{15}$,
J.~A.~M.~Lopes$^{5}$, K.~Lung$^{3}$, T.~Marrod\'an~Undagoitia$^{15,6}$, F.~V.~Massoli$^{14}$,
A.~J.~Melgarejo~Fernandez$^{1}$, Y.~Meng$^{3}$, M.~Messina$^{1}$, A.~Molinario$^{12}$, J.~Naganoma$^{16}$, K.~Ni$^{13}$, U.~Oberlack$^{7}$, 
S.~E.~A.~Orrigo$^{5}$\footnote{Present address: IFIC, CSIC-Universidad de Valencia, Valencia, Spain}, 
E.~Pantic$^{3}$, R.~Persiani$^{14}$, F.~Piastra$^{6}$, G.~Plante$^{1}$, N.~Priel$^{8}$, A.~Rizzo$^{1}$, S.~Rosendahl$^{9}$, J.~M.~F.~dos~Santos$^{5}$, G.~Sartorelli$^{14}$, J.~Schreiner$^{15}$, M.~Schumann$^{17,6}$, L.~Scotto~Lavina$^{11}$, M.~Selvi$^{14}$, P.~Shagin$^{16}$, H.~Simgen$^{15}$, A.~Teymourian$^{3}$, D.~Thers$^{11}$, O.~Vitells$^{8}$, H.~Wang$^{3}$, M.~Weber$^{15}$ and C.~Weinheimer$^{9}$\\(The XENON100 Collaboration)}

\columbia
\nikhef
\losangeles
\assergi
\coimbra
\zurich
\mainz
\weizmann
\munster
\purdue
\subatech
\torino
\shanghai
\bologna
\heidelberg
\houston
\bern

\eads{\mailto{scotto@subatech.in2p3.fr}, \mailto{jacob.lamblin@lpsc.in2p3.fr}}

\begin{abstract}
The XENON100 dark matter experiment uses liquid xenon in a time projection chamber (TPC) to measure xenon nuclear recoils resulting from the scattering of dark matter Weakly Interacting Massive Particles (WIMPs). In this paper, we report the observation of single-electron charge signals which are not related to WIMP interactions. These signals, which show the excellent sensitivity of the detector to small charge signals, are explained as being due to the photoionization of impurities in the liquid xenon and of the metal components inside the TPC. They are used as a unique calibration source to characterize the detector. We explain how we can infer crucial parameters for the XENON100 experiment: the secondary-scintillation gain, the extraction yield from the liquid to the gas phase and the electron drift velocity.
\end{abstract}
\submitto{\JPG}


\section{Introduction}
\begin{figure}[ht]
\begin{center}
\includegraphics[height=4.2cm] {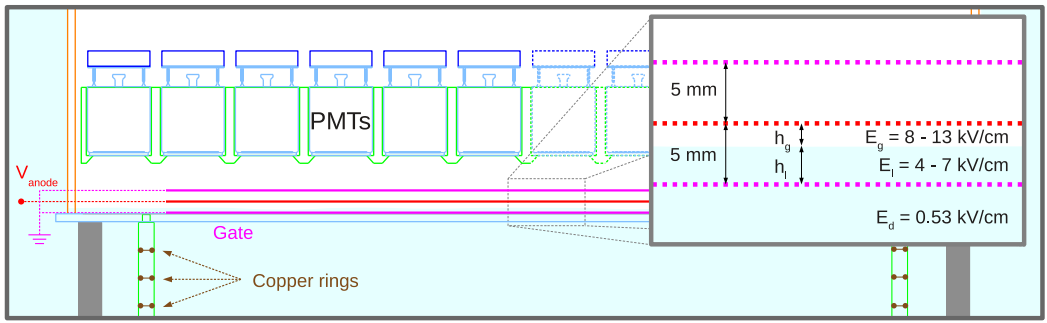}
\end{center}
\caption{Top part of the XENON100 TPC showing the top PMT array, the anode and the two grounded grids. Some of the copper rings used as field shaper along the whole TPC are also visible. Grey areas indicate the PTFE used as an insulator and reflector for the VUV scintillation light.}
\label{fig:topTPC}
\end{figure}

The XENON100 experiment~\cite{XE100INST}, which aims at the direct detection of dark matter, has been in operation at the Laboratori Nazionali del Gran Sasso (LNGS) in Italy
 since 2009. The detector is a double phase (liquid/gas) time projection chamber (TPC) filled with xenon. A WIMP interaction inside the liquid phase will produce a nuclear recoil that can be detected via the ionization and excitation of xenon atoms and molecules \cite{KUBOTA}. Excitation and recombination of some ionization electrons lead to the quasi-instantaneous emission of VUV photons ($\sim$178\,nm \cite{JORTNER1965}), which is the primary scintillation (S1). Ionization electrons that escape recombination drift toward the top of the TPC under an electric field of 530\,V/cm and eventually reach the liquid-gas interface. Electrons are extracted to the gas phase where, under a higher electric field $E_g$ ($\sim$12\,kV/cm), secondary scintillation (S2) is produced \cite{APRILEDOKE,CHEPELARAUJO}. The drift electric field in the liquid xenon volume is produced between a cathode at negative potential and a grounded gate grid. Forty field shaping rings, regularly spaced along the TPC wall, ensure the homogeneity of the field. With those boundaries, the TPC is 30.1\,cm high with a radius of 15.3\,cm. The stronger electric field $E_g$ needed for the electron extraction is obtained by means of an anode grid placed 5\,mm above the gate (Figure \ref{fig:topTPC}). A technique similar to the one of a diving bell was chosen to keep the liquid in the TPC at a precise level ($h_l$) between the gate and the anode grids. Finally, a second grounded grid is placed at 5\,mm above the anode.
Both S1 and S2 photons, corresponding respectively to light and charge signals produced by a particle interaction in the sensitive volume, are detected by 178 photomultiplier tubes (PMTs) located at the top and bottom of the TPC. The distinction between S1 and S2 signals is based on the signal duration. Because secondary scintillation photons are emitted along the path of the electrons in the gas gap $h_g$ between the liquid surface and the anode, 
S2 signals are much longer than S1 signals (S2 signal is $\sim1\,\mu$s, mainly coming from a mean velocity of few mm/$\mu$s \cite{DIAS} and a gas gap of few mm, while S1 is $\sim50$\,ns).
The $x$-$y$ position of the interaction is reconstructed from the S2 hit pattern on the top PMTs, which are arranged in concentric circles in order to achieve the best radial position reconstruction. The corresponding $z$ position is given by the time difference between the S1 and S2 signals (the drift time) multiplied by the electron drift velocity in the liquid. Calibration sources are inserted through a copper tube which is wound around the cryostat at half height of the TPC. A full description of the detector can be found in~\cite{XE100INST}.

The most recent publications, using 224.6 live days of data, provided upper limits on spin-independent~\cite{XE100RUN10PRL} and spin-dependent~\cite{XE100SD} WIMP interactions. More details on the analysis of XENON100 data can be found in~\cite{XE100ANALYSIS}.

This paper focuses on the very low-energy part of the charge spectrum (S2). In the first section, we report the observation of very low-energy S2 signals and describe their different characteristics. In the second section, we discuss the origin of these signals. Finally, we explain how we use them to characterize some aspects of the detector that are related to the ionization signal. 

\section{Observation of single-electron signals }

\begin{figure}[ht]
\begin{center}
\includegraphics[height=5.8cm] {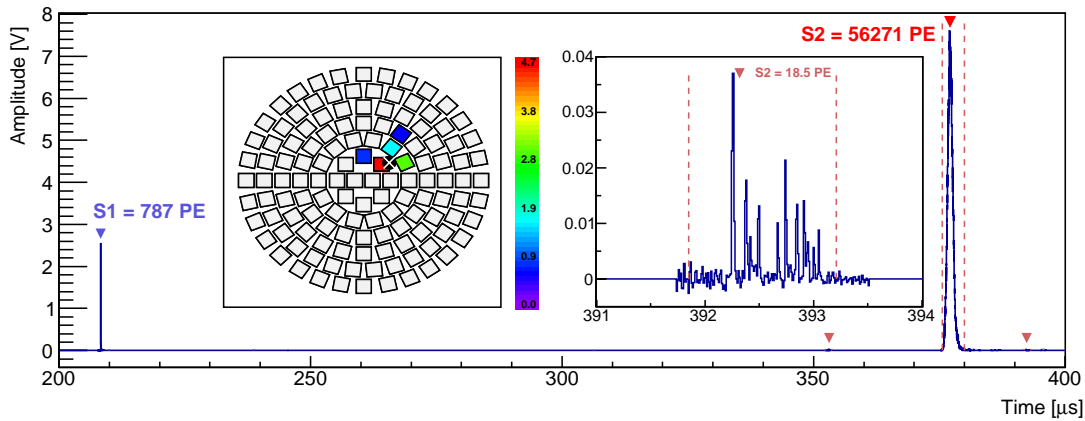}\\
\end{center}
\caption{Example of a XENON100 waveform, with the primary (S1) and secondary (S2) scintillation signals. Two small S2 signals below 150\,PE are observed and indicated by the red triangles, 145\,$\mu$s after S1 and 17\,$\mu$s after the main S2. The waveform of the second one is displayed in the inset, together with its top array PMT pattern, revealing a localised signal. The color code of the legend represents the measured signal size (in PE) seen by PMTs. The X mark indicates the reconstructed $x$-$y$ position of the interaction.}
\label{fig:S2_waveform}
\end{figure}

Single-electron signals are the smallest S2 signals that can exist. In XENON100, the S2 analysis threshold for dark matter search is set to 150 photoelectrons (PE) where the trigger efficiency is almost 100\,\%. However, lower S2 signals that do not generate trigger can be found in the $400\, \rm \mu s$ waveform of events triggered by a S1 or a S2 signal. Figure \ref{fig:S2_waveform} shows an example of waveform containing S2 signals below 150\,PE. The ZEPLIN collaboration already reported the observation of such a kind of signals (\cite{ZEPLIN2008}, \cite{ZEPLIN2011}).

\subsection{Low-energy S2 spectrum}
The first step in our study is to ensure that small S2 signals as low as $10$\,PE are real charge signals. First, their duration is around 1\,$\mu$s, which is consistent with the time needed for an electron to drift through the proportional scintillation gas gap $h_g$. This feature allows the identification of S2 signals with respect to S1 signals induced by primary scintillation, whose width is typically $\sim50$\,ns. Second, small S2 signals have a mean value for the ratio of the summed top PMT signal amplitude over the summed bottom PMT signal amplitude of 1.3. This value agrees with the value for photons emitted from the gas gap, which has been estimated both with Monte Carlo simulations and with experimental data using higher-energy events.

\begin{figure}[ht]
\begin{center}
\hspace*{-1.cm}
\includegraphics[height=6.5cm] {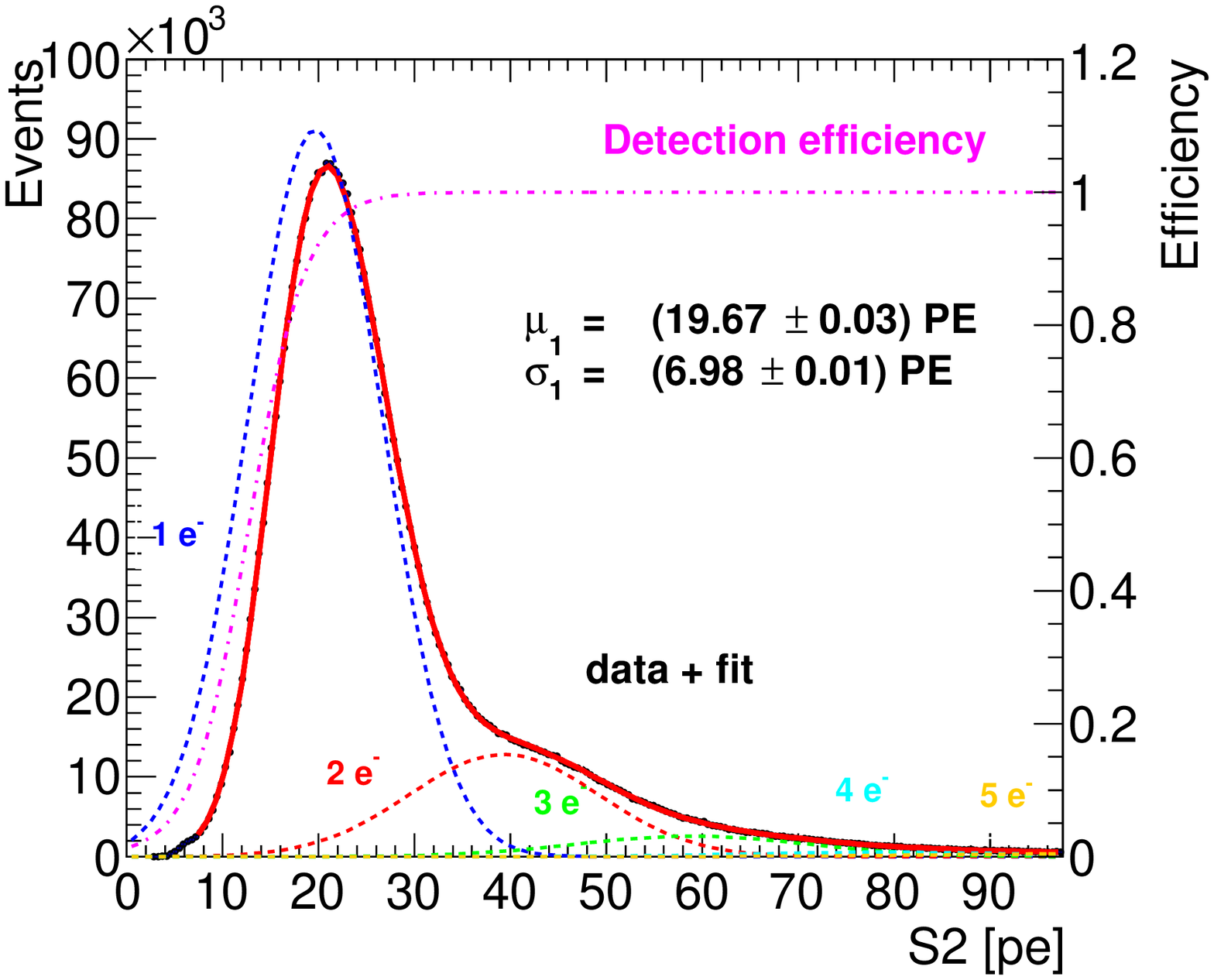}
\includegraphics[height=6.5cm] {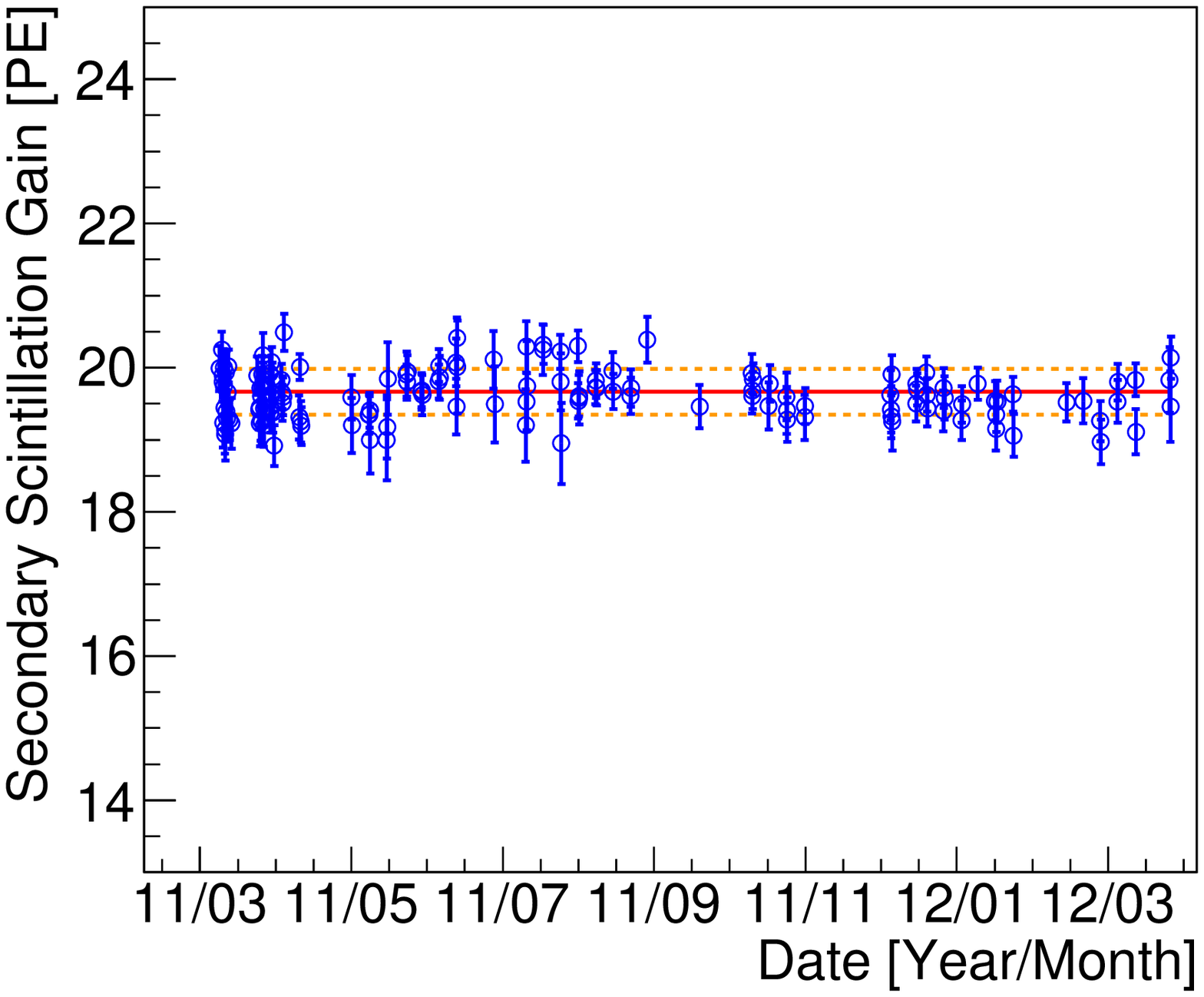}
\end{center}
\caption{(left) A typical S2 low-energy spectrum is fitted with a sum of 5 Gaussians, supposing that the spectrum comprises a sum of one to five electron's S2 signals, multiplied by a function (see text) to take into account the detection efficiency. The fit result in red superimposes the data in black. (right) Stability of the first Gaussian's mean value $\mu_1$, which defines the secondary scintillation-gain. The mean is given by the solid line and the one sigma values by the dashed lines. Periods where data are missing correspond to maintenance periods.}
\label{fig:S2_spectrum}
\end{figure}

A typical S2 low-energy spectrum is presented in Figure \ref{fig:S2_spectrum} (left).
A peak is observed at about 20\,PE and a smaller one at about 40\,PE. This observation indicates some discrete phenomenon being responsible for these small S2 signals.
The S2 spectrum shown in Figure \ref{fig:S2_spectrum}, obtained using calibration data from a $^{60}$Co source, is fitted using a sum of several Gaussian functions with mean $\mu_i$ and standard deviation $\sigma_i$ with the constraint $\mu_i = i \mu_1$ and $\sigma_i = \sqrt{i} \sigma_1$, multiplied by an efficiency curve, represented by the function $f(E) = 1/(exp(-(E-A)/B)+1)$ with $A$ and $B$ as free parameters. This fit assumes that the low-energy spectrum comprises a sum of one to a few electrons S2 signals and that each electron produces an independent S2 peak distributed as a Gaussian with mean $\mu_1$ and standard deviation $\sigma_1$. The efficiency curve is interpreted as the efficiency of the S2 peak-finder algorithm, which depends on the S2 peak size. The position of the first Gaussian ($\mu_1$) provides the number of detected photoelectrons per single electron extracted into the gas gap. This quantity is called \emph{secondary-scintillation gain}; its value depends on the physical properties of the xenon gas gap, such as the electric field, the size of the gap, and the xenon pressure. The value of $\sim20\,$PE is compatible with the secondary-scintillation gain that can be inferred from gamma calibrations by dividing the measured S2 signal by the known energy deposit and multiplying it by the effective $W$-value, i.e. the average energy expended per electron escaping recombination (see Section~\ref{sec:extract}). 

In order to check the reliability of the fit, we verified that results are constant in time under the same operational conditions (Figure \ref{fig:S2_spectrum}, right). We also checked that the position of the first Gaussian mean is the same for different calibration sources ($^{60}$Co, $^{232}$Th, $^{137}$Cs and $^{241}$AmBe) and without any source.

\subsection{Time distribution} \label{sec:TimeDistrib}
The time distribution of the small S2 signals can help to reveal time correlation with other large signals and to identify their origin.
 
\begin{figure}[ht]
\begin{center}
\hspace*{-0.5cm}
\includegraphics[height=7.5cm,angle=0] {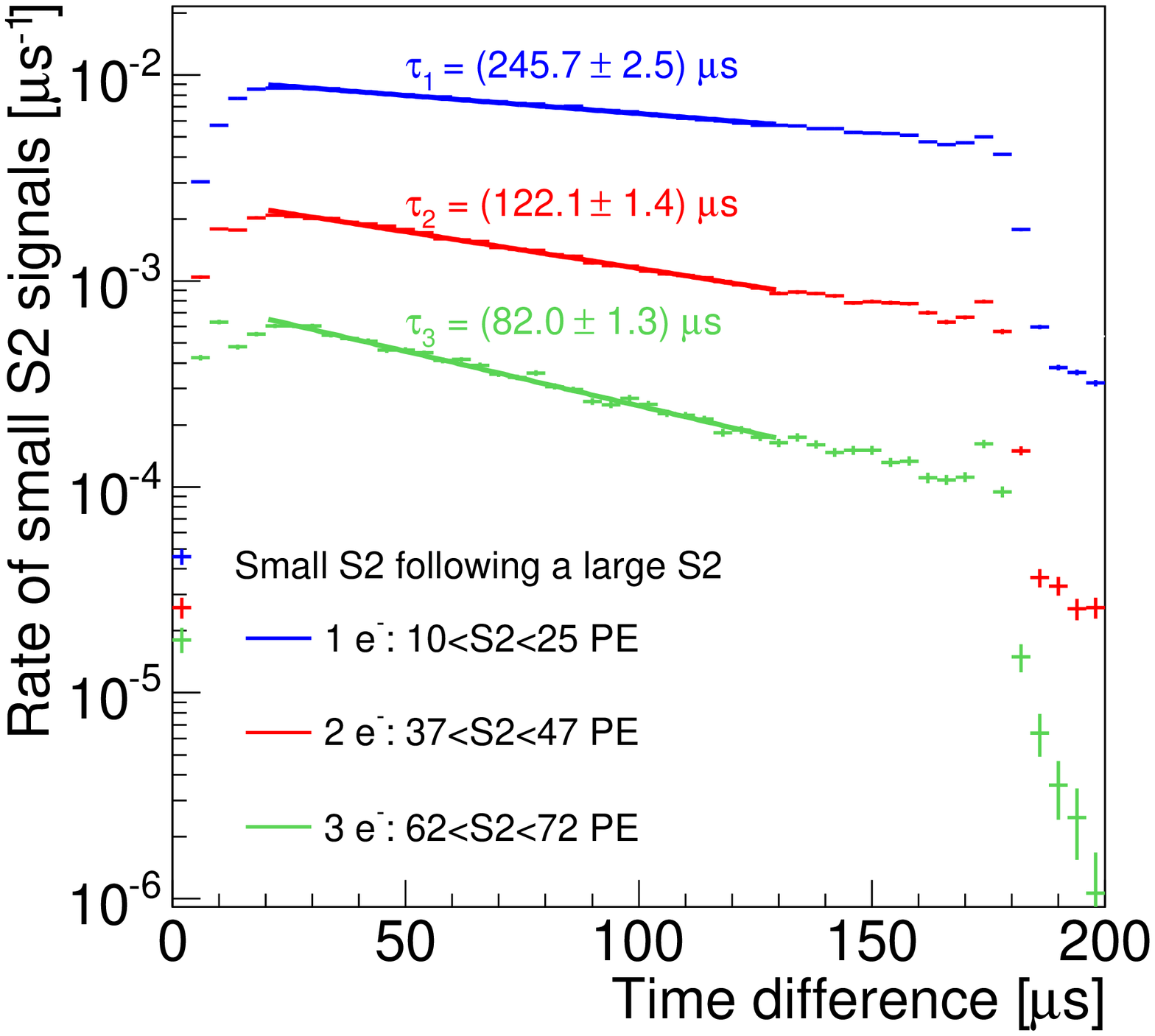} 
\hspace*{-0.5cm}
\includegraphics[height=7.5cm,angle=0] {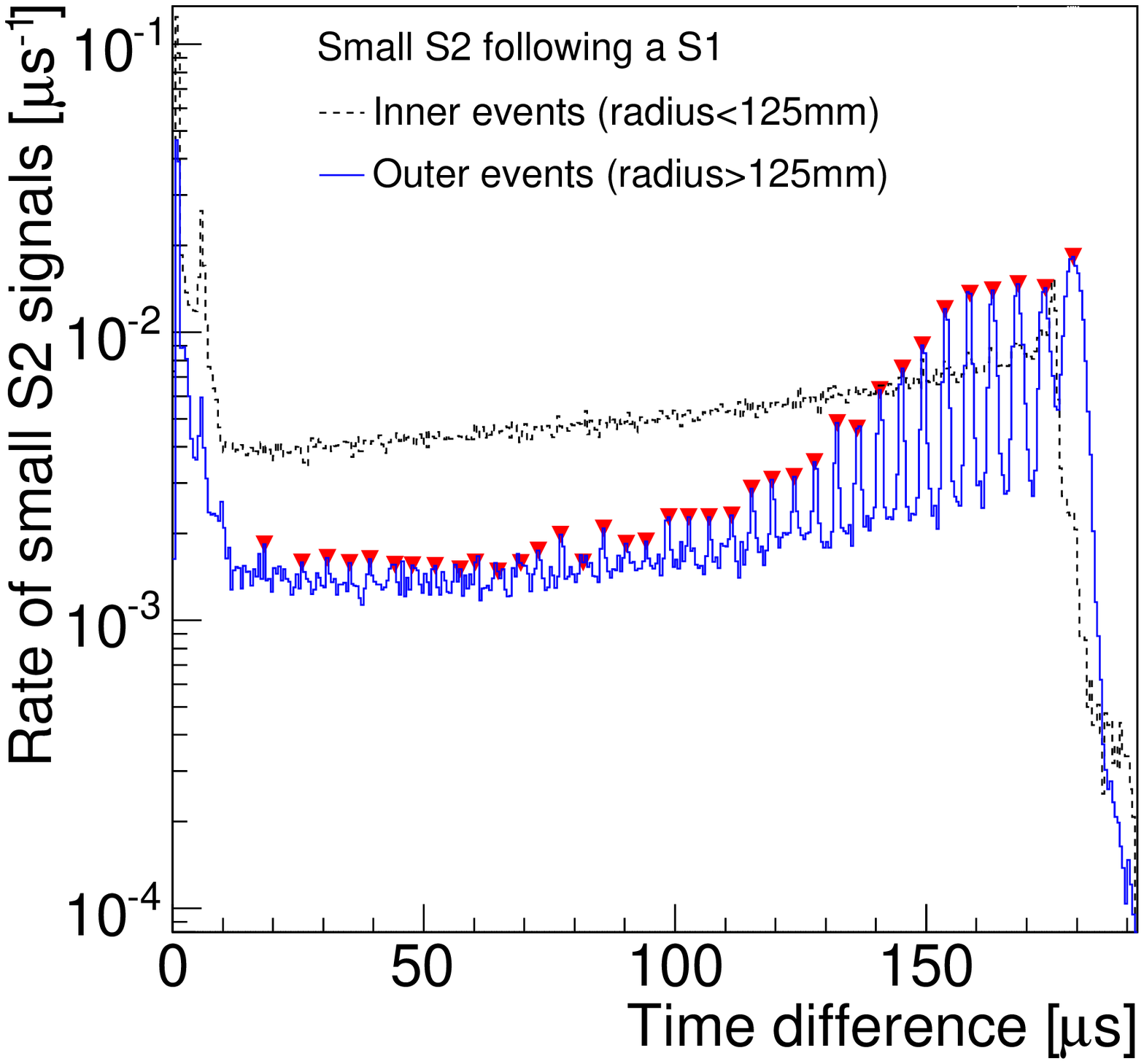} 
\end{center}
\caption{(left) Distribution of the time difference between the main S2 and the single-electron signals (blue), two-electron signals (red) and three-electron signals (green) that follow the main S2, for events with only one main S2 signal larger than 150 PE. No radial cut has been applied. Lines correspond to exponential fits and the inferred time constants are given with statistical uncertainties only. The cut-off of distributions around the maximum drift time of 180\,$\mu$s shows that the origin of small S2 signals is correlated with main S2. The time constants inferred from the fit show that multi-electron signals are accidental coincidences of single-electron signals. (right) Distribution of the time difference between the S1 ($>500$ PE) and the small S2 signals for S1-triggered events with no S2 larger than 150 PE. The distributions are divided into two radial populations. For details see text.}
\label{fig:TimeDistrib}
\end{figure}

Figure \ref{fig:TimeDistrib} (left) presents the distributions of the time difference between the main S2 signal and the small S2 signals contained in triggered event waveforms from $^{60}$Co calibration datasets. Cuts on the S2 size have been used to select single-electron, two-electron and three-electron signals. The ranges have been defined in order to select pure samples, i.e. parts of the spectrum where Gaussians do not overlap.
To avoid confusion in the time association and to study a possible time correlation, only single-scatter events containing the main S2 signal above 150\,PE have been selected. 
Because the small S2 signal rate depends on the size of the main S2 signal, as we show in Section~2.3, the $y$-axis in Figure \ref{fig:TimeDistrib} (left) corresponds only to a mean value of the rate of small S2 peaks.

The time-difference distributions show a sharp drop around 180\,$\mu$s, which corresponds to the maximum drift time in the TPC (being the maximum drift length 30\,cm divided by the drift velocity 0.173\,cm/$\mu$s measured in XENON100 \cite{XE100ANALYSIS}). This feature demonstrates that a correlation between signals exists even if some signals are still observed later than 180\,$\mu$s (see Section~\ref{sec:origins} for an interpretation). The decrease at low time differences ($<~20~\mu \rm s$) is explained by the lower efficiency of the small S2 signal detection algorithm due to the presence of several larger S2 signals coming from multiple-scatter events and, to a lesser extent, due to the width of the main S2 signal.

The distributions follow decreasing exponential functions whose time constant depends on the size of the small S2 signals. By fitting the time distributions, we find that the time constants of the two-electrons case, $(122.1\pm1.4)\,\mu$s, and three-electrons case, $(82.0\pm1.3)\,\mu$s, are respectively half and a third of the single-electron one, $(245.7\pm2.5)\,\mu$s (error bars only account for statistical uncertainties). These relationships can be explained if multi-electron signal results from accidental time coincidences of single electrons.
Indeed, if $R_{1}$ is the rate of single electrons, the accidental time coincidence of $n$ single electrons is $R_{n}=R_{1}^n \cdot \Delta t ^{n-1} $, where $\Delta t$ is the time coincidence window, which corresponds to the mean S2 width ($\sim$1$\,\mu$s). As we observe in the figure, the electron rates decrease with time by following the expression $R_{n}(t) = R_n(0) \cdot exp(-t/\tau_n)$. By substituting this equation in the previous formula we obtain the final relation $\tau_n=\tau_1/n$. 
The accidental coincidence scenario is also supported by the PMT hit pattern of the multi-electron signals, which are not localized around one PMT but rather spread over the PMT array. 

Figure \ref{fig:TimeDistrib} (right) presents the drift-time distribution of small S2 signals following a S1 signal for triggered events that do not have any S2 signal above 150\,PE. This condition allows to remove the small S2 signals that are correlated with a large S2 signal.
Single electrons are divided into two populations: 
inner events (radius $r<125$\,mm) and outer events ($r>125$\,mm), the radius of the TPC being $150$\,mm. Even if the position reconstruction algorithm is not 
optimized for such low signals, the spatial resolution is still good enough to allow the use of the reconstructed $x$-$y$ position. As for signals following a large S2 signal, the distributions end around 180\,$\mu$s.
Moreover, the outer events distribution shows regularly-spaced peaks. Red arrows indicate the position of all the peaks on the outer events distribution found using a peak finder algorithm. The mean time difference between peaks is $(4.23\pm0.05)\,\mu$s.
The interpretations of these observations are given in Section~\ref{sec:origins}.
Finally, the time distribution does not show a lower rate at small time differences as the efficiency to separate an S1 from an S2 close to each other is larger than in the case of two S2s. This is related to the shorter duration of the S1 pulse compared to the S2 one.

\subsection{Rate} \label{sec:Rate}
\begin{figure}[ht]
\begin{center}
\begin{tabular}{cc}
\hspace*{-1.cm}
\includegraphics[height=7.2cm,angle=0] {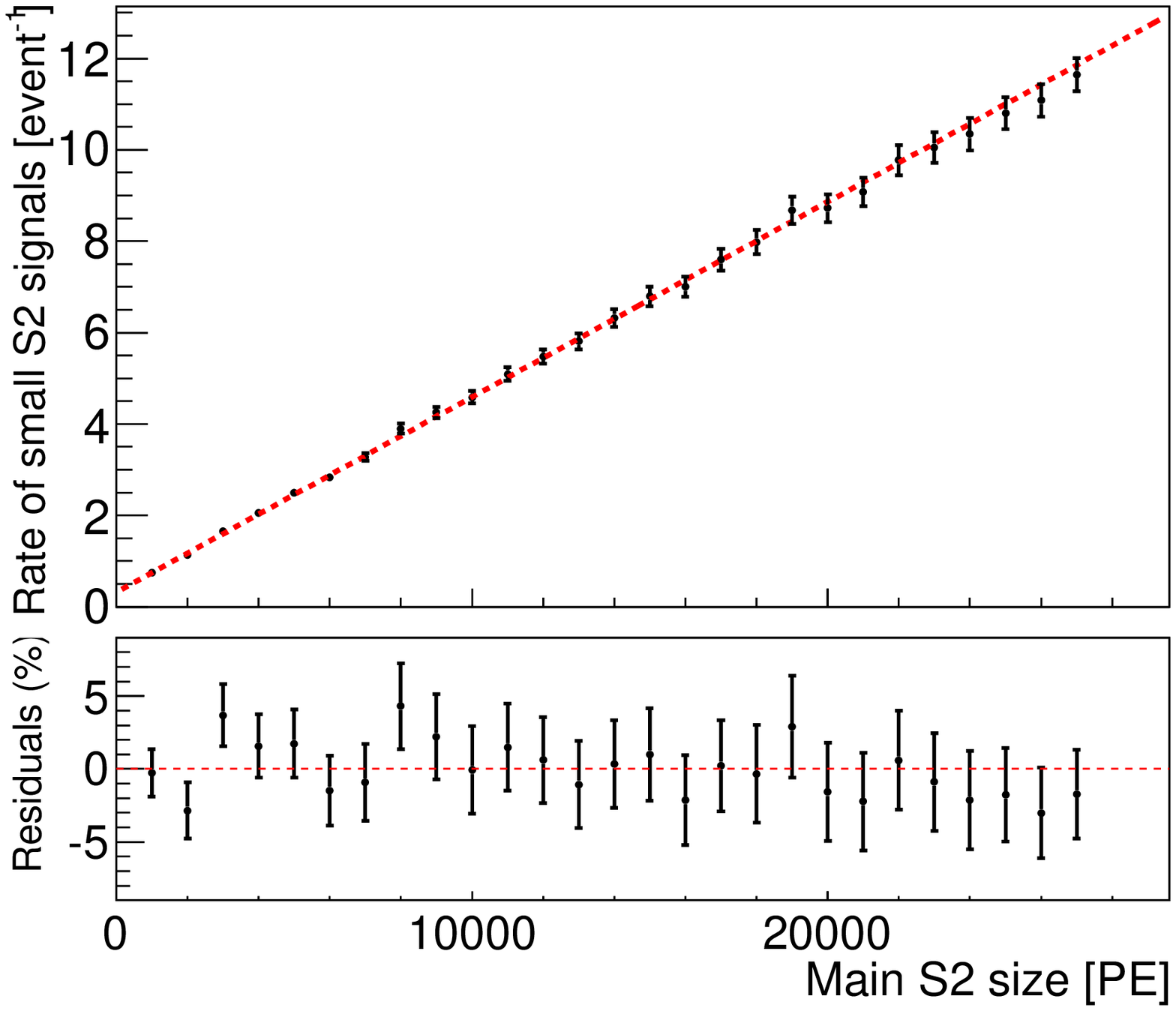}& 
\hspace*{-1.5cm}
\includegraphics[height=7.6cm,angle=0] {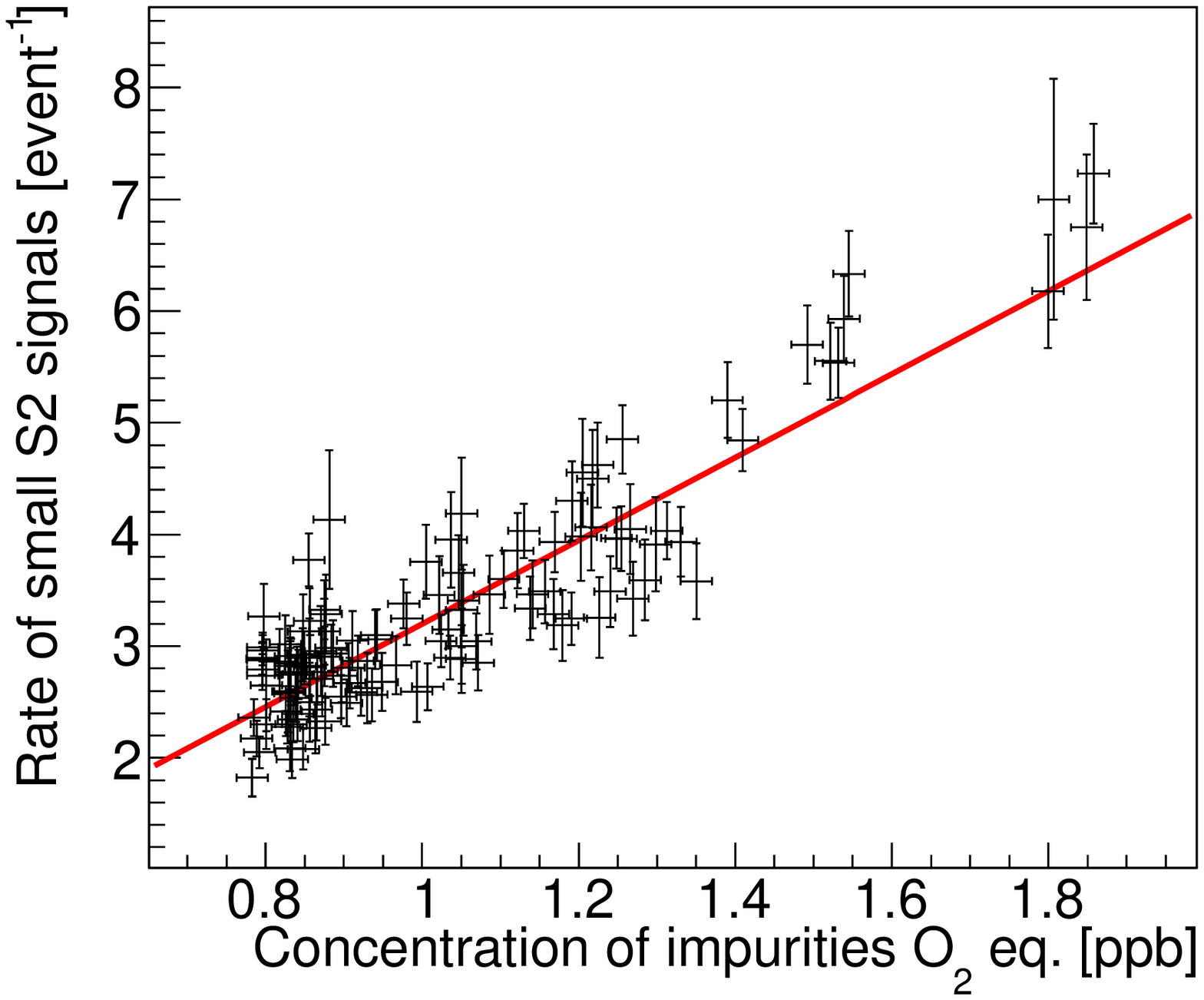} \\
\end{tabular}
\end{center}
\caption{(Left) single-electron rate per event as a function of the main S2 signal size. The residuals of the data points with respect to the linear fit show a very good proportionality of the relation. (Right) single-electron rate per event, for events with the main S2 between 5000 and 10000 PE, as a function of the O$_2$-equivalent concentration of impurities in liquid xenon. The linear fit shows that the rate is also proportional to the concentration of impurities.
}
\label{fig:SingleRate}
\end{figure}

Another interesting observable is the rate of small S2 signals.
Figure \ref{fig:SingleRate} (left) shows that the relative rate per triggered event of small S2 signals ($<150$\,PE) following the main S2 is proportional to the size of the main S2 signal. In this figure, the small S2 signals are taken from 0 to 180\,$\mu$s, the maximum drift time, after the main S2 signal. All events with just one S2 signal of more than 150\,PE size in the whole waveform of 400\,$\mu$s length are considered.

Datasets used for this figure were recorded during the last months of the dark matter search of 225 days when the purity was the highest and almost constant \cite{XE100RUN10PRL}. A linear fit gives a proportionality coefficient of $4.3\cdot10^{-4}$ small S2 signals per photoelectron in the main S2 signal and an ordinate at the origin of 0.3 small S2 signal per waveform, which corresponds to signals not correlated to the main S2 signal.

The relative single-electron rate is also proportional to the impurity concentration in the liquid xenon, as shown on Figure \ref{fig:SingleRate} (right). The figure presents the relative rate of small S2 signals that are located from 20 to 150\,$\mu$s after the main S2 signal of single-scatter events. This smaller time window has been chosen to select only the single electrons that are potentially generated inside the liquid xenon volume. 
The oxygen-equivalent impurity concentration is calculated based on the electron lifetime measured with the 662-keV photopeak from a $^{137}$Cs source \cite{XE100INST} and the O$_2$ attachment rate K$_{\rm {O}_2}$, which is $9.7\cdot10^{10}\,$l/mol/s at 0.53\,kV/cm and 182K, obtained by linearly interpolating the values at 87K and 165K taken from \cite{SCHMIDT}. 
For this figure, $^{137}$Cs source calibration datasets recorded over a period of 18 months have been used. The analysis has been repeated using background data (without any external source), leading to the same conclusions. 

We also observe single-electron signals not associated with a S1 or a S2 signal. To estimate the corresponding rate, we looked at the part of the waveform before the trigger for which there is no S1 or S2 signal.
We collected the equivalent of about 100\,s of waveform per day. We derived, in case of absence of an external source, a relative rate of $5\cdot10^{-3}$ fortuitous single electrons per triggered event, which is much lower than the rate of single electrons associated with a S1 or a S2 signal. In addition, the relative rate exponentially decreases, with a time constant of 2\,ms, when the time delay since the previous triggered event increases. 

\section{Origin of single-electron signals}
\label{sec:origins}

There are several possible origins to the single-electron signals. While some of them would result from the photoionisation of impurities in the liquid and photoelectric effect on the stainless steel of the cathode, the others could be a consequence of delayed extraction of electrons at the liquid-gas interface or of field emission at the cathode. From all observations reported in the previous section, we can confirm some of these conclusions.

First, most of the small S2 signals are induced by primary or secondary scintillation photons. This is the only way to explain the time distributions (Figure \ref{fig:TimeDistrib}).
Delayed extraction of electrons at the liquid-gas interface (i.e., electrons which would be trapped at the liquid surface and extracted later than the main part of the electron cloud) could explain the exponential decrease for single electrons following the S2 signal but not the sudden end at a time corresponding to the maximum drift time.

VUV photons can induce electrons by the photoelectric effect, and one photon can only generate one electron. This is compatible with our observations since all small S2 signals are single electrons or accidental coincidences of single electrons, as we have shown in Section~\ref{sec:TimeDistrib}.
Within this scenario, the shape of the time distributions would result from the convolution of the photon spatial distribution and the exponential attenuation due to drift electron attachments. If non uniform, the spatial distribution of photoionisation targets would also play a role.
Thus, the different shapes of the time distributions in Figure \ref{fig:TimeDistrib} are a consequence of the photon's origin.
Photons from the charge signal S2 are emitted from the top of the TPC, which explains the decreasing shape of the time distributions of single electrons induced by S2.

For the S1 induced events (Figure \ref{fig:TimeDistrib}, right), we select S1-only events which are mainly from interactions below the cathode. Then, more single electrons are generated at the bottom of the TPC, i.e. at larger drift times.

The photoelectric effect explains also the proportionality between the relative rate of small S2 signals and the main S2 size, i.e. the number of secondary photons, as shown in Figure \ref{fig:SingleRate}. From the proportionality coefficient derived in this figure and taking into account the mean light collection efficiency in XENON100 for photons emitted from the gas gap ($\bar{\beta}\sim20\,\%$ from Monte Carlo simulation) and the averaged PMT quantum efficiency ($\sim25\,\%$) \cite{XE100INST}, we conclude that about 50\,000 secondary scintillation photons are needed on average to produce one single electron with impurities molecules in the xenon at the ppb level. The non-zero ordinate at the origin, corresponding to an average of $0.32\pm0.02$ additional single electrons per triggered event, corresponds to single electrons that are either induced by the S1 photons or not induced by any photon.
 
There are several candidates for being the target for the photoionisation observed in the TPC: xenon, impurity molecules (O$_2$, N$_2$, ...) 
contained in the xenon at the ppb level or the detector components (grid, cathode, field shaping rings,...). The correlation of the rate of small S2 signals with impurity concentration shown in Figure 5 suggests that the dominant photoionization process in the XENON100 detector is on the impurity 
molecules in liquid xenon. Given the VUV photon energy of $\sim7.0$\,eV, the negative O$_2$ ions created by the attachment of drift electrons are the best candidates because the needed energy is $\sim0.45$\,eV \cite{O2AFFINITY}, while the first ionization energy of O$_2$ and N$_2$ are above 12\,eV (\cite{O2IONIZATION}, \cite{N2IONIZATION}). However, since the cross sections and the number of ions are unknown, 
it is not possible to make any quantitative statement. We cannot exclude photoionization of other chemical species.

The photoionization of impurities is not the only cause of single electrons. This can be deduced from the peaks shown in Figure \ref{fig:TimeDistrib} (right). The peaks can be explained by photoelectric effects on the copper of the forty field shaping rings and on the stainless steel of the cathode. In terms of drift distance, the mean time difference between peaks corresponds to $(7.32\pm0.09)$\,mm, which agrees with the separation of the field shaping rings, $(7.15\pm0.01)$\,mm. The peaks with higher time differences are much larger for two reasons. First, for this analysis, primary photons inducing single electrons are emitted from the bottom of the TPC, as explained above. Second, due to the non uniformity of the field lines at the bottom of the TPC at large radii \cite{XE100INST}, the electrons emitted from the bottom field shaping rings drift toward the center of the TPC and reach the anode more easily than the electrons emitted from upper field shaping rings. 
The large event number in the first bins of both distributions comes from photoelectric effect on the stainless steel of the gate grid, which separates the drift volume from the amplification gap and is located $5$\,mm below the anode (see Figure \ref{fig:topTPC}).

Finally, the fortuitous low-energy S2 signals observed without S1 or S2 signal and the ones that arrive later than the maximum drift time cannot be directly induced by primary or secondary photons. The observed time correlation with the previous triggered event at the millisecond scale suggests that a delayed extraction phenomenon occurs.

Also the ZEPLIN collaboration suggested several origins to the single-electron signals (\cite{ZEPLIN2008}, \cite{ZEPLIN2011}) and our results are consistent with what they found.

\section{Detector characterization using single electrons}

Single electrons are a unique calibration source to characterize the detector's performance related to the ionization process. In this section, we present detailed analyses that lead to the measurement of the electron extraction yield from the liquid to the gas, the secondary-scintillation gain, and the electron drift velocity in the liquid.

\subsection{Secondary scintillation gain} \label{sec:SecScintGain}
The secondary-scintillation gain $G$ is defined as the total number of photoelectrons, observed by all the PMTs in the TPC, per electron extracted into the gas gap. It is related to the secondary scintillation yield $Y$, i.e. the number of emitted photons per electron extracted in the gas gap, by the expression:

\begin{equation}
G(E_g,P_g,h_g) = Y(E_g,P_g,h_g)~\bar{\beta}~\bar{\eta},
\label{EqG}
\end{equation}
where $\bar{\beta}$ is the mean collection efficiency of photons emitted from the gas gap, and $\bar{\eta}$ is the averaged product of PMT quantum and photocathode collection efficiencies. The quantity $Y$ depends on the properties of the gas gap where the photon emission occurs: the electric field $E_g$, the pressure $P_g$, and the height $h_g$ of the gap. It is usually described by:

\begin{equation}
Y= (a \frac{E_g}{P_g} + b )~h_g~P_g,
\label{EqY}
\end{equation}
where $a$ and $b$ are parameters that have been measured under several conditions (see \cite{MONTEIRO} for a compilation of existing measurements and simulation results).

\begin{figure}[ht]
\begin{center}
\includegraphics[height=8.cm] {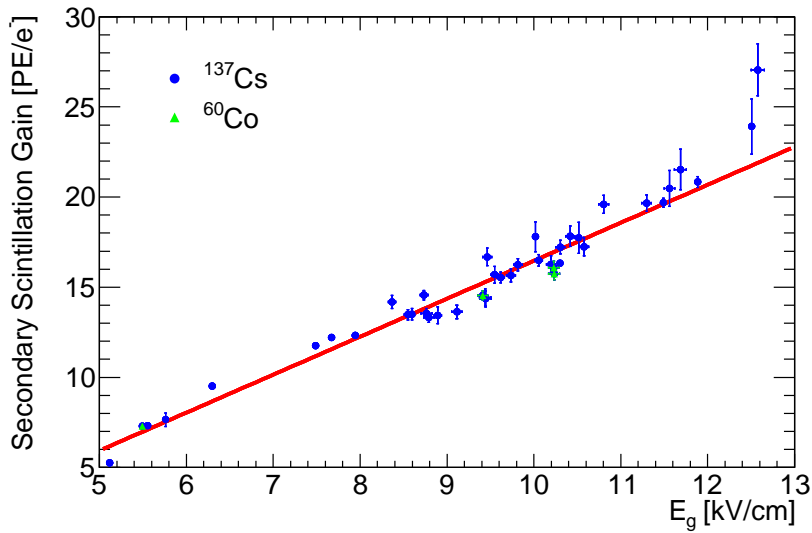}
\end{center}
\caption{XENON100 secondary-scintillation gain as a function of the electric field in the gas gap. The secondary-scintillation gain is proportional to the electric field. The two highest values do not follow the linear trend due to the onset of electron multiplication in the gas gap.
}
\label{fig:SecScintGain}
\end{figure}

The secondary-scintillation gain can be obtained from the mean size of single electron S2 signals, corresponding to $\mu_1$, the mean value of the first Gaussian in the fit of the low-energy S2 spectrum, which has been described in Section~2.1. Figure \ref{fig:SecScintGain} presents the secondary-scintillation gain as a function of the electric field using calibration data ($^{137}$Cs and $^{60}$Co) recorded at different anode voltages $V_a$ (from 2.2\,kV to 4.5\,kV) and gas gap heights $h_g$ (from 1.3\,mm to 4.1\,mm), leading to an electric field ranging between 5.25\,kV/cm and 12.55\,kV/cm. The electric field is calculated from the relation $E_g=\epsilon_r V_a/(\epsilon_r h_g + d-h_g)$ where $d$ is the distance between the gate grid and the anode ($d=5$\,mm) and $\epsilon_r=1.96$ \cite{DIELECTRIC} is the dielectric constant of liquid xenon. The pressure was very stable at 2.248\,bar during the acquisition time, with fluctuations $<0.24\,\%$ \cite{XE100INST}.

Since $G$ linearly depends on $h_g$, all gain values have been rescaled to the same gas gap height used as a reference ($h_g=2.9$\,mm) to present results in a consistent way. A fit with the function (\ref{EqG}), with $a$ and $b$ as free parameters, yields $a=(151\pm19)$\,photons/e$^-$/kV and $b=-(147\pm19)$\,photons/e$^-$/cm/bar. The uncertainties are dominated by the uncertainty on the collection efficiencies. These values are in good agreement with those presented in \cite{MONTEIRO} and predicted by Monte Carlo simulation \cite{SANTOS} for room temperature and also with those observed for saturated xenon vapour at cryogenic temperatures \cite{FONSECA}. The highest values of the secondary-scintillation gain that do not follow the linear trend are affected by electron multiplication as the electric field is here above the ionization threshold.

For the two XENON100 dark matter search runs of 100.9 live days \cite{XE100RUN8PRL} and 224.6 live days \cite{XE100RUN10PRL} and for the run started at the end of 2012, the secondary-scintillation gains have been estimated using all calibration sources and the dark matter search datasets. We obtain, as average value, $(18.7\pm0.7)$\,PE/e$^-$, $(19.7\pm0.3)$\,PE/e$^-$ and $(17.1\pm0.4)$\,PE/e$^-$, respectively and $(6.6\pm0.7)$\,PE/e$^-$, $(6.9\pm0.3)$\,PE/e$^-$ and $(6.4\pm0.2)$\,PE/e$^-$ as standard deviation (corresponding to $\sigma_1$ of Section~2.1). The errors take into account also systematic uncertainties. The difference between the mean values comes from different gas gap heights and anode voltages.

\subsection{Electron extraction yield} \label{sec:extract}
\begin{figure}[ht]
\begin{center}
\includegraphics[height=8.cm] {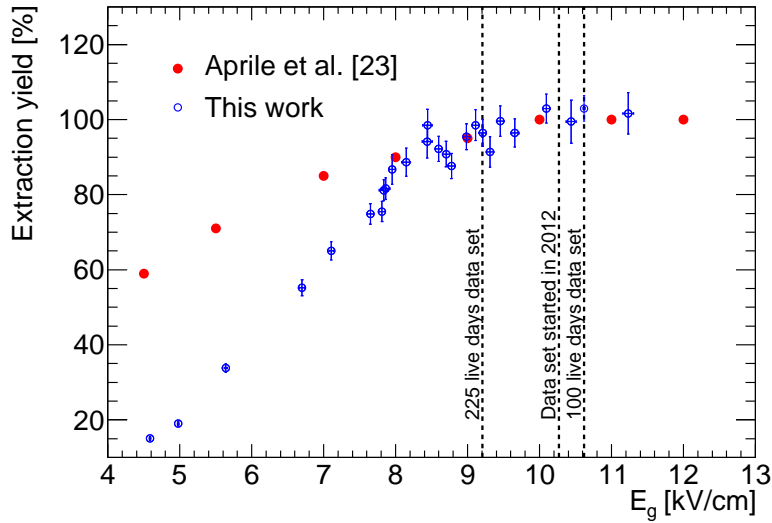} 
\end{center}
\caption{XENON100 liquid-gas extraction yield as a function of the electric field in the gas gap, calculated at the surface of the liquid. Vertical lines represent the field strength during the two dark matter search runs already published and the one started at the end of 2012. In both cases, the extraction yield was close to unity. For comparison, the data points of \cite{APRILEIEEE} are shown in red. For the fields relevant for XENON100, the results are in agreement.}
\label{fig:ExtractionYield}
\end{figure}

The extraction yield is the probability for an electron to be extracted from the liquid phase into the gas phase.
It is an important parameter since it affects the S2 resolution, as the S2 resolution depends primarily on the number of transmitted electrons to the gas phase.
The extraction yield can be obtained by dividing the number $N_g$ of electrons that are extracted into the gas phase by the number $N_l$ of electrons that reached the liquid-gas surface. To extract this ratio, we selected electronic recoils from full absorption of 662\,keV gammas emitted by a $^{137}$Cs source. The quantity $N_l$ is obtained by dividing 662\,keV by the mean energy needed to produce an electron-ion pair ($W=15.6\pm0.3$\,eV \cite{TAKAHASHIPRA}), scaled to the fraction of electrons which do not recombine with positive ions, $T_{ee}$, and corrected for the electron lifetime of our data. The quantity $N_g$ is inferred from the data by selecting the full-absorption peak in the S2 spectrum and by dividing the observed mean value by the secondary-scintillation gain $G$.
The extraction yield inferred from this method is presented in Figure \ref{fig:ExtractionYield} as a function of the electric field in the gas gap. The field is calculated at the surface of the liquid since this field is the one which is responsible for the electron extraction into the gas phase.

As expected, the extraction yield increases with the electric field, reaching a plateau around $10$\,kV/cm.  
Below 8\,kV/cm, the current measurement gives lower values than the published ones~\cite{APRILEIEEE}. The reason of the discrepancy is unknown and it might be explained by the different geometrical configuration or the different estimation method.
Requiring the plateau to be at $100\,\%$ yield, we inferred from our data an effective $W$-value (i.e., the $W$-value divided by $T_{ee}$ to correct for the recombination) of $(23.5\pm0.7)\,$eV. Considering the published $W$-value, it corresponds to a fraction of electrons which do not recombine, $T_{ee}=0.66\pm0.02$, for an electric field of 530\,V/cm. This value is in agreement with the published value ($T_{ee}=0.74\pm0.07$ at 662\,keV at an electric field of 0.5\,kV/cm \cite{APRILEPRB}) but with a better precision. This result is also in agreement with the prediction of the NEST model \cite{NEST,SZYDAGIS}, in which the value $T_{ee}=0.63$ (with 4\% of systematic error estimated) is derived for the same electric field.

The extraction yield for the XENON100 dark matter search runs are close to unity at 10.6\,kV/cm for the 100 days data set \cite{XE100RUN8PRL}, 10.2\,kV/cm for the run started at the end of 2012 and 9.2\,kV/cm for the 225 days data set \cite{XE100RUN10PRL}.

\subsection{Liquid level and drift velocity measurement} \label{sec:liqlevel}
\begin{figure}[ht]
\begin{center}
\begin{tabular}{cc}
\hspace*{-1.cm}
\includegraphics[height=7.cm,angle=0] {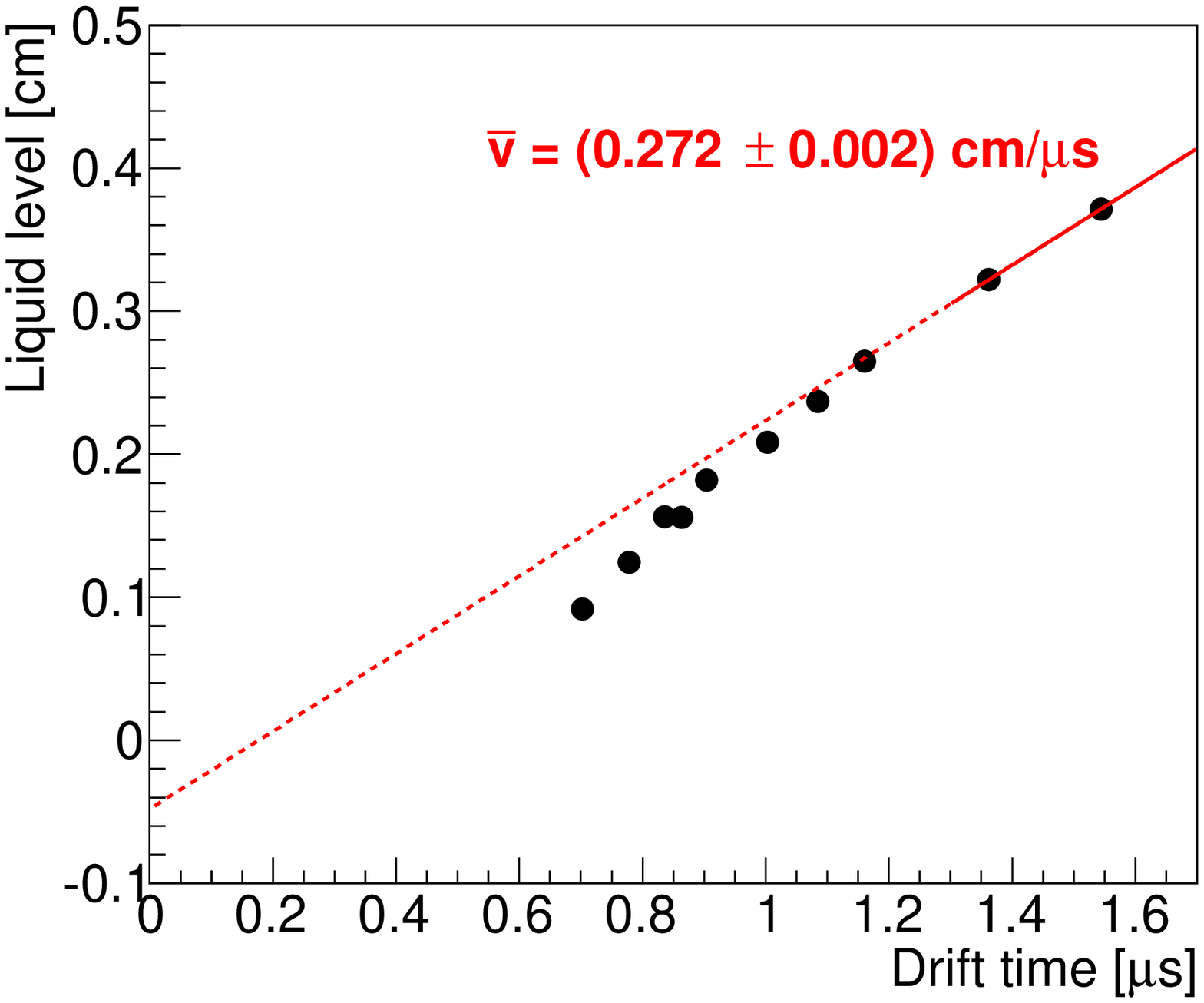}
\hspace*{-0.5cm}
\includegraphics[height=7.cm,angle=0] {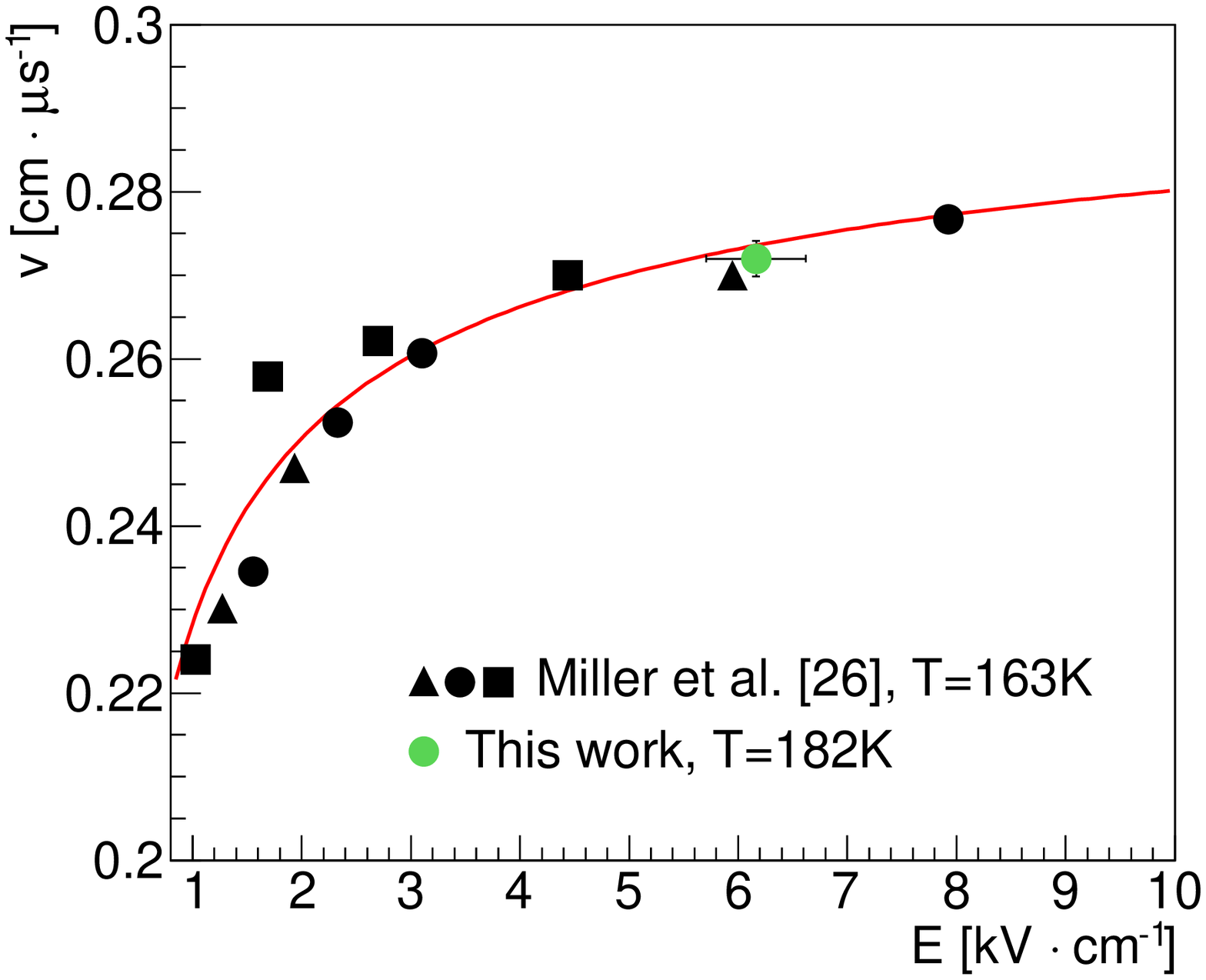} 
\end{tabular}
\end{center}
\caption{(left) Average time interval between S1 and S2 for single electrons emitted from the gate grid as a function of the liquid level. The error bars are within the symbols. The non-linearity could be explained by a bias on the time estimation or on the non-uniformity of the electric field. The last two points are used to estimate the drift velocity. (right) Drift velocity measured in XENON100 in an electric field of $(6.2\pm0.5)$\,kV/cm (green circle). The measured drift velocity is in perfect agreement with published values.
The black points and the curve are the measurements obtained in \cite{MILLER} and the $E^{1/2}$ dependency derived from this data (markers indicate different specimens as described in the referred paper).}
\label{fig:LiquidLevel}
\end{figure}

As shown in Section~2, the detector is sensitive to single electrons emitted from the gate grid due to the photoelectric effect on the stainless steel by primary scintillation photons. For these events, the time difference between the primary scintillation and the single-electron signal corresponds to the drift time for electrons between the gate grid and the liquid-gas interface where the secondary scintillation starts. Figure \ref{fig:LiquidLevel} (left) shows the mean drift time of single electrons from the gate grid for different values of the liquid level from 0.9\,mm to 3.7\,mm above the gate grid, measured with an anode voltage of 4\,kV. The liquid level is determined using a capacitive level meter \cite{XE100INST} with a relative precision of 80\,$\mu$m. The mean drift time is inferred from a gaussian fit of the time difference peak selecting all events with S2 below 150\,PE.

The relation between liquid level and drift time is expected to be linear at high liquid levels because of the small dependency of the drift velocity on the electric field range, $E=(4-7)$\,kV/cm, present between the gate grid and the liquid surface, as shown in ~\cite{MILLER}. The non-linearity observed at low liquid levels can be explained by the minimum time difference between peaks that the peak-finder algorithm can resolve ($\sim0.6\,\mu$s), which leads to an overestimation of the mean time difference. A non-uniform electric field near the grounded gate grid would give also such an effect. 

We can infer the drift velocity in the liquid xenon by fitting the experimental points in the region at high liquid levels, where we make the assumption that the linear domain is reached. As we do not have enough data points, we limit our fit to the last two points and use the third point to determine the systematic error. For an electric field of $(6.2\pm0.5)$\,kV/cm, corresponding to an anode voltage of 4\,kV and a liquid level of 0.34\,cm, the measured drift velocity is $\bar{v}=(0.272\pm0.002)$\,cm/$\mu$s.
Without systematic error, the ordinate at the origin of the fit should be zero. 
The non-zero ordinate ($\sim 0.5$\,mm) gives an estimation of the absolute error of the level meter, which is expected to be the dominant one. Thus, single electrons provide a way to improve the liquid level measurement.
 
Figure \ref{fig:LiquidLevel} (right) shows the field dependence of the drift velocity presented by Miller {\it et al.}~\cite{MILLER} for liquid xenon at 163\,K. The solid curve was obtained by fitting their experimental results (represented by markers indicating different specimens as described in the referred paper) with the function $E^{1/2}$ for values of electric fields greater than $E=0.1$\,kV/cm. As stated in Ref.~\cite{MILLER}, a complete theoretical model describing such a dependence at higher fields is still missing. Our measured value agrees with the expectation from literature, confirming the validity of the method.

\section{Conclusions}
We have reported the observation of very low energy secondary scintillation pulses in the XENON100 dark matter dual-phase TPC. We have demonstrated that these signals are caused by single electrons extracted into the gas phase or accidental coincidences of those single electrons. The events are mainly generated by the photoionization of impurities in the liquid and of metal components (copper field shaping electrodes and stainless steel grids inside the TPC), by primary or secondary scintillation VUV photons. These single-electron signals have been used to study TPC characteristics such as secondary scintillation gain, electron extraction yield into the gas phase, liquid xenon level, and electron drift velocity. The results obtained using single electrons are in good agreement with those obtained using other methods or in literature.

\ack
We gratefully acknowledge support from NSF, DOE, SNF, Volkswagen Foundation, FCT, Region des Pays de la Loire, STCSM, NSFC, DFG, MPG, Stichting voor Fundamenteel Onderzoek der Materie (FOM), the Weizmann Institute of Science, the EMG research center and INFN. We are grateful to LNGS for hosting and supporting XENON100.


\section*{References}
\begin{thebibliography}{9}

\bibitem{XE100INST}
E. Aprile {\it et al.} (XENON100 Collaboration),
Astropart. Phys. 35, 573 (2012).

\bibitem{KUBOTA}
S. Kubota {\it et al.}, 
Phys. Rev. B 17, 2762 (1978). 

\bibitem{JORTNER1965}
J. Jortner, L. Meyer, S. A. Rice, and E. G. Wilson, 
J. Chem. Phys. 42, 4250 (1965).

\bibitem{APRILEDOKE}
  E.~Aprile and T.~Doke,
  Rev.\ Mod.\ Phys.\  82, 2053 (2010).

\bibitem{CHEPELARAUJO}
  V.~Chepel and H.~Araujo,
  JINST 8, R04001 (2013).

\bibitem{DIAS}
T.H.V.T. Dias {\it et al.},
Phys. Rev. A 48, 2887 (1993).

\bibitem{XE100RUN10PRL}
E. Aprile {\it et al.} (XENON100 Collaboration), 
Phys. Rev. Lett. 109, 181301 (2012).

\bibitem{XE100SD}
E. Aprile {\it et al.} (XENON100 Collaboration), 
Phys. Rev. Lett. 111, 021301 (2013).

\bibitem{XE100ANALYSIS}
E. Aprile {\it et al.} (XENON100 Collaboration), 
Astropart. Phys. 54, 11 (2014).

\bibitem{ZEPLIN2008}
B. Edwards {\it et al.}, 
Astropart. Phys. 30, 54 (2008).

\bibitem{ZEPLIN2011}
E. Santos {\it et al.}, 
JHEP 2011, 115 (2011). 

\bibitem{SCHMIDT}
G. Bakale, U. Sowada, W.F. Schmidt, 
J. Phys. Chem., 80, 2556 (1976).

\bibitem{O2AFFINITY}
M.J. Travers, D.C. Cowles, G.B. Ellison, 
Chem. Phys. Lett. 164, 449 (1989).

\bibitem{O2IONIZATION}
R.G. Tonkyn, J.W. Winniczek, M.G. White,
Chem. Phys. Lett. 164, 137, (1989).

\bibitem{N2IONIZATION}
T. Trickl, E.F. Cromwell, Y.T. Lee, A.H. Kung, 
J. Chem. Phys., 91, 6006, (1989).

\bibitem{MONTEIRO}
C.M.B. Monteiro et al, 
JINST 2, P05001 (2007).

 \bibitem{DIELECTRIC}
W.F. Schmidt, 
Proceedings of the International Workshop on Technique and Application of Xenon Detectors, Kashiwa, Japan,
World Scientific, (2003) 1.

\bibitem{SANTOS}
F.P. Santos {\it et al.}, 
J. Phys. D: Appl. Phys. 27, 42 (1994).

\bibitem{FONSECA}
A.C. Fonseca {\it et al.},
2004 IEEE Nucl. Sci. Symp. Conference Record, 1, 572


\bibitem{XE100RUN8PRL}
E. Aprile {\it et al.} (XENON100 Collaboration), 
Phys. Rev. Lett. 107, 131302 (2011).

\bibitem{TAKAHASHIPRA} 
T. Takahashi {\it et al.},
Phys. Rev. A 12, 1771 (1975).

\bibitem{APRILEPRB}
E. Aprile et al, 
Phys. Rev. B 76, 014115 (2007).

\bibitem{APRILEIEEE}
E. Aprile {\it et al.}, 
IEEE Trans. Nucl. Sci. 51, 5 (2004).	

\bibitem{NEST}
  M.~Szydagis, N.~Barry, K.~Kazkaz, J.~Mock, D.~Stolp, M.~Sweany, M.~Tripathi and S.~Uvarov {\it et al.},
  JINST 6, P10002 (2011).

\bibitem{SZYDAGIS}
  M.~Szydagis, A.~Fyhrie, D.~Thorngren and M.~Tripathi,
  JINST 8, C10003 (2013).


\bibitem{MILLER}
L.S. Miller, S. Howe, W.E. Spear,
Phys. Rev. 166, 871 (1968).

\end{thebibliography}
\end{document}